# COMPREHENSIVE COMPLEXITY ASSESSMENT OF EMERGING LEARNED IMAGE COMPRESSION ON CPU AND GPU

*Farhad Pakdaman[1,2] and Moncef Gabbouj[1]*

[1]Faculty of Information Technology and Communication Sciences, Tampere University, Finland
[2]Faculty of Engineering and Technology, University of Mazandaran, Babolsar 4741613534, Iran

**ABSTRACT**

Learned Compression (LC) is the emerging technology for compressing image and video content, using deep neural networks. Despite being new, LC methods have already gained a compression efficiency comparable to state-of-the-art image compression, such as HEVC or even VVC. However, the existing solutions often require a huge computational complexity, which discourages their adoption in international standards or products. This paper provides a comprehensive complexity assessment of several notable methods, that shed light on the matter, and guide the future development of this field by presenting key findings. To do so, six existing methods have been evaluated for both encoding and decoding, on CPU and GPU platforms. Various aspects of complexity such as the overall complexity, share of each coding module, number of operations, number of parameters, most demanding GPU kernels, and memory requirements have been measured and compared on Kodak dataset. The reported results (1) quantify the complexity of LC methods, (2) fairly compare different methods, and (3) a major contribution of the work is identifying and quantifying the key factors affecting the complexity.

## 1. INTRODUCTION

As image and video content continue to dominate internet traffic, new compression techniques are developed to cope with their high bitrate. Emerging multimedia applications such as UHD video streaming, virtual reality, and cloud gaming, are among the demanding applications that require a superior compression efficiency for practical deployment. Learned Compression (LC) is the relatively new solution that deploys Deep Neural Networks (DNN) for image and video compression [1][2][3][4]. These methods either use DNN modules as parts of the compression pipeline, i.e., learned coding tools [5], or train a DNN end-to-end to serve as the encoder-decoder pair [2][3]. The latter approach is known as the end-to-end LC, and is widely investigated in both academia and industry.

While different solutions can be very different, Fig. 1 shows a generic framework for the end-to-end learned image compression. The image is transformed by an analysis network, $g_a$, to form a latent space. The entropy Arithmetic Encoder (AE) generates the bitstream which is transmitted to the receiver for decoding. After Arithmetic Decoding (AD), the image is reconstructed from the latent, using the synthesis transform network $g_s$. To find the best entropy parameters that account for latent space dependencies, often a pair of auxiliary networks ($h_a$ and $h_s$) are used to transform the latent space into a compact hyperprior. This information, sometimes alongside other context information is used to derive the entropy parameters.

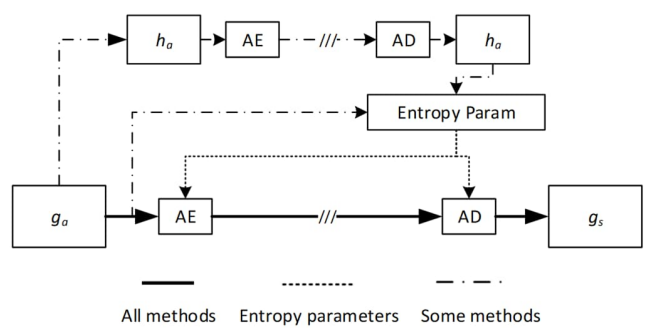

Fig. 1 a simplified visualization of LC image codec

Ballé et al. [2] proposed one of the first successful solutions, which uses an autoencoder architecture for $g_a$ and $g_s$. First a simple method with a factorized prior (**FP**) is introduced. Then a hyperprior (**HP**) is proposed that considers the dependency in latent space. Minnen et al. [3] improved this model by generalizing it to a gaussian mixture model, with mean and scale values conditioned on the prior (**MS-HP**). Authors also propose an autoregressive context model (**ARC**), which serially models the entropy based on seen latent information. Cheng et al. [4] proposed to use a Discretized Gaussian Mixture Likelihood, to model the latent information (**DGM**). They also studied the use of attention mechanism in encoder and decoder networks, to enhance the performance (**DGM-ATT**). Several other solutions have been proposed that introduce novel techniques for image and video compression. Self-organized variational autoencoders [6], techniques to achieve flexible rate points [7], and task-oriented compression [8] are among these efforts. However, the six highlighted methods above are among the most reputable, and are the basis for most other methods.

Successful LC proposals have also encouraged multimedia standardization activities. JPEG-AI [9] is the upcoming image compression standard based on end-to-end LC, with the goal of improving compression efficiency for human vision, and also enabling latent (compressed) domain machine vision. MPEG NNVC [10] and MPEG VCM [11] are also two upcoming video standards that are heavily based on LC.

Although existing LC-based solutions succeed in enhancing compression efficiency, their huge computational complexity have so far been a big obstacle in their standardization, and prevented them to be deployed in products. This becomes more worrisome when considering that multimedia applications are already complex and reported to constitute 1% of the global carbon emission [12]. Moreover, the encoding cost of state-of-the-art is already so high, that even streaming giants only consider using them for their most popular content [13]. Complexity analysis and profiling [14][15] have been performed for previous standards, which encouraged significant contributions in reducing the encoding search space or efficient implementations [16]. This is even more crucial for the emerging LC technology, as (1) despite traditional methods, their encoding/decoding process are rather deterministic, with no search space to optimize, and (2) unlike traditional standards where decoding is kept much simpler than encoding, LC decoding is comparable to its huge encoding complexity. Hence, it is even more important to consider the complexity requirements of proposed techniques before standardization or deployment.

With this motivation, this paper presents a thorough complexity assessment of several existing LC methods. To do so, six of the most popular methods have been selected to analyze on Kodak dataset [17]. An evaluation methodology has been used to analyze both encoding and decoding operations on CPU and GPU platforms. Using Nvidia NSight System [18] and precise manual annotations, the following aspects have been measured: (1) total encoding and decoding times were measured and compared with both LC and traditional codecs, (2) important sub-modules' share of total time were measured, (3) the number of parameters and number of kilo Multiply and Accumulate (kMAC) operations were measured for each sub-module, (4) Hotspot CUDA kernels were identified and their share of complexity were quantified, and finally (5) memory transactions required for each method were measured. We believe these detailed measurements and observation can play a crucial role in deployment of LC-based technology, and can be an important guideline for its future development.

The remainder of the paper is organized as follows. Section 2 details the methodology used for complexity assessment. Section 3 summarizes the detailed experimental results, and finally Section 4 concludes the paper. As it is not possible to include all detailed experimental results in the paper, more detailed results and data can be found in the project webpage https://github.com/farhad02/LC_Assessment.

## 2. COMPLEXITY ASSESSMENT METHODOLOGY

While LC methods are mainly implemented and tested on high-end GPUs, (1) GPUs are not available on all multimedia-capable devices, and (2) GPU times are difficult to be compared, especially with traditional methods. Hence, we measure both CPU and GPU complexities, to account for various scenarios. As Fig. 2 shows, each measurement is preceded by a warm-up stage which (1) loads the codec model into the device, to exclude the loading time (0.03-0.4 s, depending on model size) from the encoding/decoding times (based on JPEG-AI recommendation), (2) runs a dummy encoding/decoding to warm-up the GPU to high working frequencies, to avoid artificially large time measurements, and (3) calls for synchronization between the host and device, to avoid asynchronous execution and more accurate GPU time measurement. Next, the encoding/decoding are performed for each image on a codec fully annotated for sub-module measurements. Time measurements are done using Nsight System and Time library, CUDA kernels and memory operations are tracked via Nsight Systems, and various kernels are grouped with same methodology as [19]. Finally, the PTFlops library [20] is used to measure the number of parameters and number of kMAC operations per pixel.

These steps are repeated for encoding and decoding of each image, on both CPU and GPU platforms, for six LC methods, each on various Quality levels (Q). LC methods are FP, HP, MS-HP, and ARC, each with 8 Qs, DGM and DGM-ATT, each with 6 Qs. We use the CompressAI library [21] to run and evaluate these methods. The CPU is an Intel Core i7 11850H with a max frequency of 2.5 GHz, and the GPU is an Nvidia T1200, with 4 GBs of GPU memory and max frequency of 855 MHz.

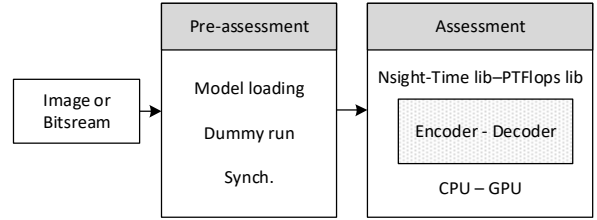

Fig 2. Complexity assessment methodology

## 4. EXPERIMENTAL RESULTS

This section summarizes and discusses the results. For the ease of read, results are summarized into figures of average numbers. More detailed tables will be released on the project webpage.

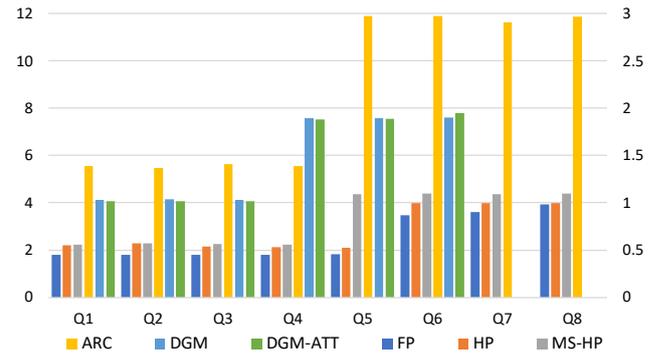

Fig 3. Total Enc times (s) on CPU. ARC, DGM, DGM-ATT on left, and FP, HP, and MS-HP on right axis

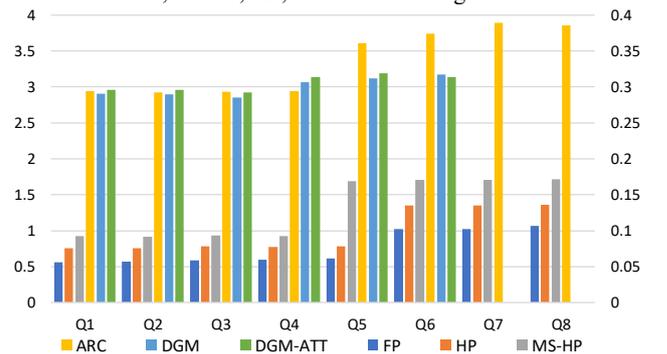

Fig 4. Total Enc times (s) GPU

**Total encoding and complexity**

Figures 3 to 6 summarize the total encoding and decoding (average of all images), on CPU and GPU. ARC, DGM, DGM-ATT are reflected on the left, and FP, HP, and MS-HP on the right axis. It is observed that for all methods complexity increases for higher Qs (better qualities). Specifically, a sudden increase is noticed on Q5 for MS-HP and ARC, on Q4 for DGM and DGM-ATT, and on Q6 for the rest. The reason is that an increased number of latent channels, and hence more activations, are used for higher quality levels (often 192, instead of 128). Almost for all figures, FP, HP, MS-HP, ARC, DGM, and DGM-ATT achieve the lowest to highest complexities. The complexity difference among different methods is huge and up to 12x and 50x for CPU and GPU encoding, and 13x and 26x for CPU and GPU decoding, which is due to different layers and different context modeling structures. Another important observation from comparing these figures is that methods show different parallelizability on GPU. First, encoding gains a better speedup on GPU compared to decoding (4.6x vs 2.2x, respectively, averaged on all methods). Second, methods with less dependency in context modeling, i.e., FA, HP, MS-HP, gain on average 8.1x, 7.1x, and 6.2x encoding speedup, while serial nature of entropy in ARC, DGM, and DGM-ATT limits them to 2.5x, 1.9x, and 1.9x speedup, respectively. Similar effect is observed for decoding.

Table II. kMAC/Pel and million parameters for each method

| | kMAC/Pel | | | | | | | | | | Million Param | | | | | | | | | |
|---|---|---|---|---|---|---|---|---|---|---|---|---|---|---|---|---|---|---|---|---|
| | Enc+Dec | | $g_a$ | | $h_a$ | | $h_s$ | | $g_s$ | | Enc+Dec | | $g_a$ | | $h_a$ | | $h_s$ | | $g_s$ | |
| Model | $Q_{high}$ | $Q_{low}$ | $Q_{high}$ | $Q_{low}$ | $Q_{high}$ | $Q_{low}$ | $Q_{high}$ | $Q_{low}$ | $Q_{high}$ | $Q_{low}$ | $Q_{high}$ | $Q_{low}$ | $Q_{high}$ | $Q_{low}$ | $Q_{high}$ | $Q_{low}$ | $Q_{high}$ | $Q_{low}$ | $Q_{high}$ | $Q_{low}$ |
| FP | 184 | 406.9 | 36.8 | 81.6 | 0 | 0 | 0 | 0 | 147.2 | 326.4 | 2.9 | 6.8 | 1.4 | 3.4 | 0 | 0 | 0 | 0 | 1.4 | 3.4 |
| HP | 188.2 | 417.1 | 36.8 | 81.6 | 1.4 | 3.3 | 2.9 | 6.7 | 147.2 | 326.4 | 5 | 11.6 | 1.4 | 3.4 | 1 | 2.4 | 1 | 2.4 | 1.4 | 3.4 |
| MS-HP | 195.3 | 437.4 | 36.8 | 81.6 | 1.4 | 3.3 | 9.9 | 27.3 | 147.2 | 326.4 | 6.9 | 17.3 | 1.4 | 3.4 | 1 | 2.4 | 3 | 8.1 | 1.4 | 3.4 |
| ARC | 412 | 450.1 | 79.2 | 81.6 | 2.4 | 3.3 | 10.2 | 27.3 | 316.8 | 326.4 | 12.1 | 20.1 | 2.8 | 3.4 | 2.2 | 2.4 | 3.3 | 8.1 | 2.8 | 3.4 |
| DGM | 412 | 925.7 | 159.5 | 358.6 | 1.5 | 3.3 | 3.4 | 7.7 | 246.5 | 552.1 | 10.9 | 24.5 | 1.8 | 4.1 | 0.7 | 1.7 | 2.7 | 6.1 | 5.2 | 11.6 |
| DGM-ATT | 457.8 | 1024.9 | 181.8 | 409.4 | 1.5 | 3.3 | 3.4 | 7.7 | 268.8 | 602.3 | 12.3 | 27.6 | 2.5 | 5.6 | 0.7 | 1.7 | 2.7 | 6.1 | 5.9 | 13.2 |

Table I. Average HEVC and VVC times (s) on CPU

| | Enc | Dec |
|---|---|---|
| AVG HM (HEVC) | 1.53 (std 0.23) | 0.06 (std 0.006) |
| AVG VTM (VVC) | 57.4 (std 17.03) | 0.09 (std 0.01) |

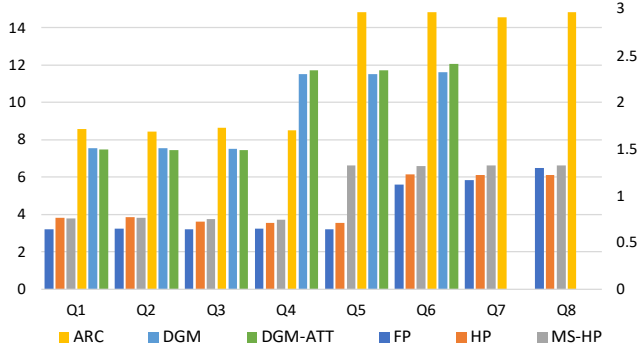

Fig 5. Total Dec times (s) CPU

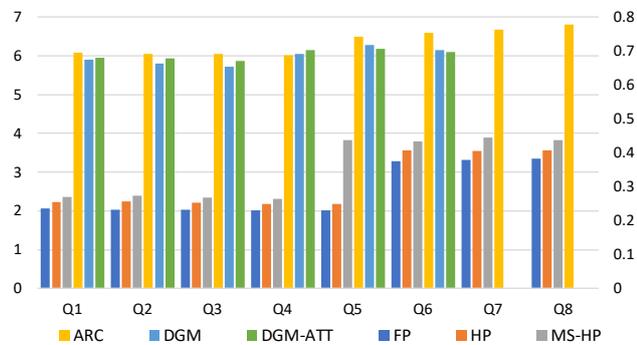

Fig 6. Total Dec times (s) GPU

To compare with traditional methods, same images were encoded and decoded with HEVC – HM 16.2 [22], and VVC – VTM 17.2 [23], on CPU (no GPU implementations are publicly available), and the average results are reported in Table I. Even thou LC methods use multiple CPU cores, comparing Fig. 3 with the single threaded implementation of HM, it is observed that most methods take a much higher time to encode. For decoding (Fig. 5), the complexity is even higher and LC methods are tens to hundreds of times slower. Even with GPU acceleration (Fig. 6) LC methods are from 4x up to 100x more complex than the CPU-only HM. It should be highlighted that despite HM and VTM, in all LC methods the decoding time is similar or comparable to its encoding time, which is undesirable for consumer devices. We also observed that for LC methods, Enc/Dec times are very similar for all images, and the coefficient of variation (STD/mean) is 0.01 on average. Compared to the large coefficient of variation for HM and VTM (0.1 to 0.3), it can be concluded that LC models are much less content dependent.

**Complexity of coding modules**

Fig. 7 reports the average share of each important coding module for each method, and for both CPU and GPU encoding/decoding. It can be observed that in CPU implementations the analysis and synthesis transforms ($g_a$ and $g_s$) take the major complexity in most methods. However, as they are highly parallelizable, they take minor times on GPU. AE and AD are the next most complex modules for encoding and decoding, and their contribution increase on GPU, as other modules are accelerated but entropy coding has more dependencies.

While execution time can depend on the execution platform, the number of parameters and operations provide a general and device-independent view of complexity. This information is summarized in Table II, for Kodak images. The lightest and heaviest codecs wr.t the number of total parameters are FP (2.9-6.8M) and DGM-ATT (12.3-27.6M), and w.r.t kMAC/Pel are FP (184-407) and DGM-ATT (458-1025), respectively. As a simplified comparison, a high-end desktop processor with 40 GFLOPs is capable of only ~100 kMAC/Pel in a second, for a similar image size.

**GPU execution details**

As DNN-based methods are often deployed on GPU platforms, further investigations were done to quantify the share of most time-consuming functions (i.e., CUDA Kernels) used in each LC method. All methods call several functions from the Nvidia cuDNN Library, which are grouped into the four main categories of Batch Normalization, Convolution, Elementwise operations, General Matrix Multiply (GeMM), and ReLU, similar to [19]. As Fig 8 shows, ReLU and convolution dominate simpler encoders, while element-wise operations dominate the more complex ones. For decoding, convolution is the dominant in three methods while for the rest, elementwise, GeMM, and ReLU constitute the majority.

As in all modern computing environments, the cost of memory transfer between the host and the GPU device is a great source of latency and energy consumption. To account for this, the memory requirements (Host to Device, HtoD, Device to Host, DtoH, Device to Device, DtoD and memset) for each method are reported in Fig 9. It is observed that (1) high Qs correspond to higher memory requirements (between 1.7 to 2.2x), (2) the autoregressive method and attention-based models are the most demanding in terms of memory requirements, and (3) decoding requires a memory similar to the encoding operation (0.65x to 0.92x).

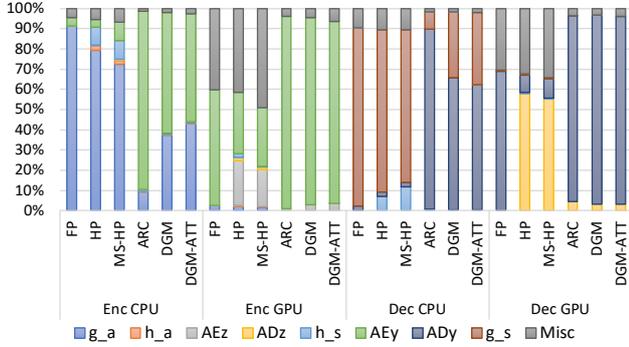

Fig 7. Module time shares for both CPU and GPU, Enc and Dec

**Rate-Distortion performance**
Although measuring the rate-distortion (R-D) performance is not an objective of this research, we provide the average R-D results over the first five images of Kodak dataset, in Fig. 10. This helps to better compare the LC methods. It can be observed that DGM and DGM-ATT methods achieve the best results. The MS-HP, ARC, and HP models achieve comparable performances, especially in low BPP, and the FP method achieves the lowest performance.

## 4. CONCLUSION

This paper presented a thorough complexity assessment of learned compression techniques, on both CPU and GPU. Six existing solutions were evaluated on Kodak dataset. It was observed that:
- The overall complexity of LC methods is much higher than traditional (HEVC and VVC). Unlike the traditional, decoding complexity of LC methods is close to their encoding complexity.
- Methods with a more complex context modeling have a significantly higher complexity at both encoding and decoding.
- Unlike the traditional, complexity of LC methods is almost content independent.
- LC methods have different parallelizability on GPU. Dec acceleration is lower than enc for all methods. Methods with high dependency in context modeling gain a limited speedup.
- Analysis transform and entropy coding are the most demanding encoding operations on CPU and GPU, respectively.
- Synthesis transform and entropy decoding are the most demanding decoding operations on CPU and GPU, respectively.
- On GPU, ReLU and convolution are the most used kernels for simpler methods, and elementwise operations and GeMM are for the more complex ones.
- Memory transfer from host to device is the largest memory usage, which almost doubles for the higher end of quality levels.

**ACKNOWLEGMENT**
This project has received funding from the European Union's Horizon 2020 research and innovation programme under the Marie Skłodowska-Curie grant agreement No [101022466].

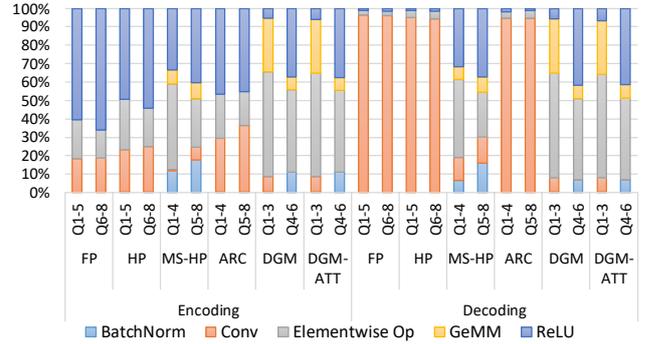

Fig 8. CUDA kernel shares for Enc and Dec

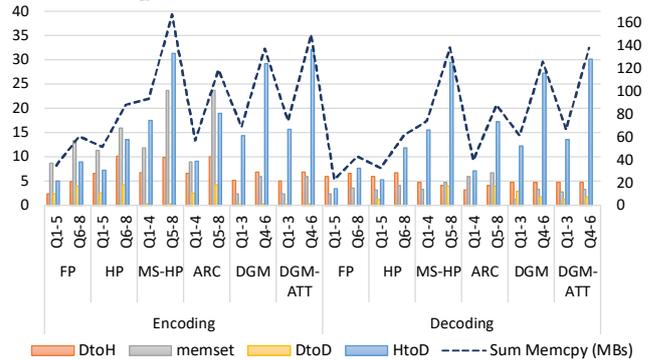

Fig 9. Memory requirements (MBs) for Enc and Dec. Sum memory and HtoD correspond to the right axis

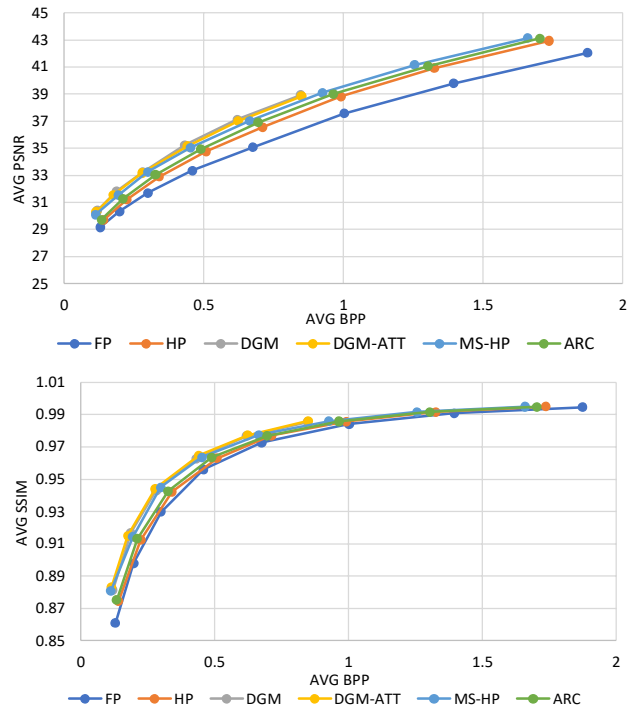

Fig 10. Average rate-distortion performance of LC methods